\documentclass[aps,prl,twocolumn,showpacs,superscriptaddress,groupedaddress]{revtex4}  
\usepackage{graphicx}  
\usepackage{dcolumn}   
\usepackage{bm}        
\usepackage{amssymb}   
\usepackage{amsmath}
\begin{document}
\title{Search efficiency of biased migration towards stationary or moving targets
in heterogeneously structured environments}
\author{Youness Azimzade}
\affiliation{Department of Physics, University of Tehran, Tehran 14395-547, Iran}
\affiliation{Leiden Academic Centre for Drug Research, Faculty of Mathematics and Natural Sciences, Leiden University, Leiden, The Netherlands} 
\author{ Alireza Mashaghi}
\email[Corresponding Author:~]{a.mashaghi.tabari@lacdr.leidenuniv.nl} 
\affiliation{Leiden Academic Centre for Drug Research, Faculty of Mathematics and Natural Sciences, Leiden University, Leiden, The Netherlands}
\date{\today}
\begin{abstract}
Efficient search acts as a strong selective force in biological systems ranging from cellular populations to predator-prey systems. The search processes commonly involve finding a stationary or mobile target within a heterogeneously structured environment where obstacles limit migration. An open generic question is whether random or directionally biased motions or a combination of both provide an optimal search efficiency and how that depends on the motility and density of targets and obstacles. To address this question, we develop a simple model that involves a random walker searching for its targets in a heterogeneous medium of bond percolation square lattice and used mean first passage time (MFPT,  $\langle T \rangle$) as an indication of average search time. Our analysis reveals a dual effect of directional bias on the minimum value of $\langle T \rangle$. For a homogeneous medium, directionality always decreases  $\langle T \rangle$ and a pure directional migration (a ballistic motion) serves as the optimized strategy; while for a heterogeneous environment, we find that the optimized strategy involves a combination of directed and random migrations. The relative contribution of these modes is determined by the density of obstacles and motility of targets. Existence of randomness and motility of targets add to the efficiency of search. Our study reveals generic and simple rules that govern search efficiency. Our findings might find application in a number of areas including immunology, cell biology, and ecology. 

\end{abstract}
\maketitle

\textit{Introduction:} 
Migration and search are ubiquitous processes in biology \cite{hein2016natural} and generic principles may underlie these processes in seemingly distinct biological contexts. Many biological organisms and objects detect the spatial gradients of chemicals and respond to them by biased migration. Examples include mice navigation for odor sources \cite{gire2016mice}, olfactory navigation in Drosophila \cite{gaudry2012smelling}, gradient sensing by amoebae and neutrophils \cite{levchenko2002models}, among others \cite{stocker2012marine, berdahl2013emergent}. Biological cells and organisms may also show diffusive motions. During immune response, T cells need to detect cognate antigens at the surface of antigen-presenting cells and to interact with other immune cells. T cells accomplish these tasks through migration in lymphatic nodes or peripheral tissue\cite{krummel2016t}. T cell migration has been reported to be diffusive \cite{miller2002two, miller2003autonomous}, sub-diffusive \cite{worbs2007ccr7, krummel2016t},  super-diffusive\cite{harris2012generalized} or a combined migration \cite{fricke2016persistence}. However, it is not clear whether these different migration modes are internally controlled or are determined by environmental conditions \cite{krummel2016t}. 

How do target properties and the environment affect the search efficiency of a cell or organism? Search processes often happen in complex environments. In peripheral tissues, cells face heterogeneous extracellular matrix (ECM) and other cells \cite{zhou2017bystander} during their migration. ECM topography is able to guide cellular migration and regulate cellular motility through physical cues that geometrically constrain adhesion sites \cite{lammermann2008rapid, petrie2009random}. For example, T cells velocity fluctuations has been attributed to morphology of lymphatic nodes \cite{beltman2007lymph} and the orientation of ECM fibrils affects direction of cellular migration \cite{wood1988contact, webb1995guidance, gomez2007polarization, teixeira2003epithelial,loesberg2007threshold, salmon2012matrix}. From such examples, one can conclude that the physical structure of the environment is a determinant of search efficiency; yet, the dependency of the optimal mode of migration on the environmental heterogeneity is not fully understood. Further, we do not know how motility and density of targets affect the optimal search strategy. For example, the effect of motility of and frequency of antigen-presenting cells (APCs) as T cell targets on the T cell search efficiency has not been studied \cite{krummel2016t}. 

Here, we use simple models to look for generic rules that determine search efficiency in a heterogeneous environment and its dependency on density and motility of targets. Random motions in the absence or presence of obstacles have been studied widely in physics literature \cite{metzler2000random}. Presence of obstacles alters both the diffusion constant \cite{nicolau2007sources, coscoy2007statistical} and the dynamics of random motion \cite{gefen1983anomalous,rammal1983random,rammal1984random} in a manner that depends on the density and structure of obstacles. Obstacles are expected to interfere with directionally biased motions as well, but the extent of this interference has not been explored for non-critical densities \textbf{ \cite{fribergh2014phase, sapozhnikov2017random, fribergh2010speed, berger2003speed, daryaei2014loop, miller2015random, havlin1987diffusion, barlow2004random}}. Here, we address this problem through simulation and quantify the efficiency using a key quantity that we borrow from studies on stochastic processes, namely first passage time (FPT, $T$, also  known as first hitting time)\cite{redner2001guide}. For an object searching for its target, FPT is a time that takes to reach the target for the first time. Mean first passage time (MFPT, $\langle T \rangle$) which quantifies the average time needed to reach a specific target, has been commonly considered as an indicative of search efficiency \cite{benichou2005optimal,benichou2011intermittent,sheinman2012classes}. MFPT has been calculated for both Markovian and non-Markovian walkers \cite{condamin2005first, condamin2007first, benichou2010geometry, benichou2014first, guerin2016mean}, but it has not been studied (neither analytically nor through simulation) for biased random walk in the absence or presence of obstacles. Furthermore, the effect of motility of targets (with both random and biased migration) and the effect of obstacles on this process has not been studied yet.  As we explain in the following, our findings reveals generic rules that might be applicable to a wide range of problems including immune cell migration, where search efficiency is being intensely researched and MFTP has not yet been used as a measure of search efficiency \cite{ krummel2016t, harris2012generalized}. 
 
\begin{figure}  [htb]
 	\centerline{\includegraphics[width=0.5\linewidth]{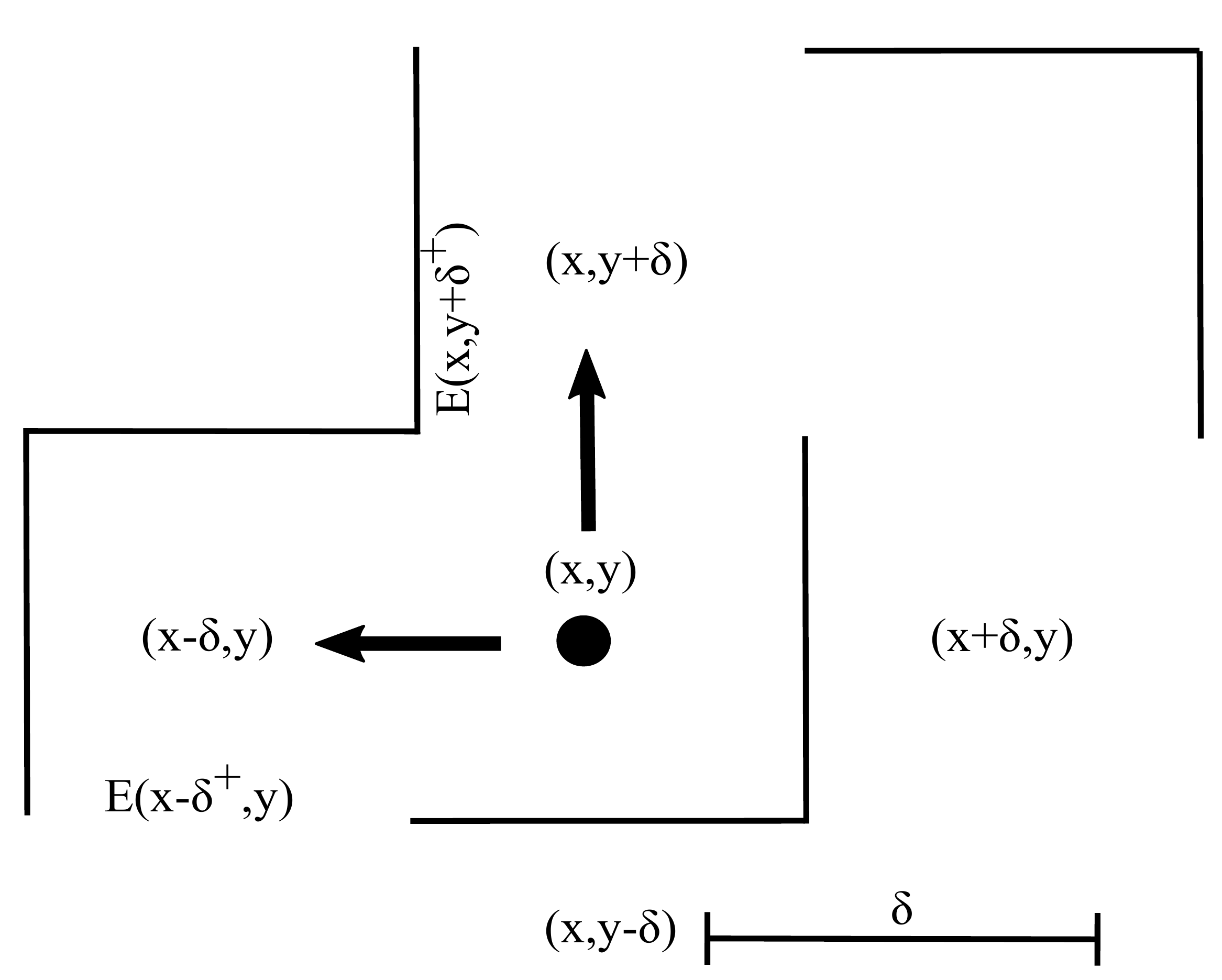}}
 	\caption{Illustration of a searching walker within a structured medium. The walker cannot jump to the neighboring units separated by a solid line (corresponding $E$ is equal to one), while it is free to jump otherwise (corresponding $E$ is equal to zero). As such, in this figure, the walker at $(x,y)$ is only allowed to move in depicted directions. As an initial condition, each line will be considered solid with probability of $p$ (This process creates the square bond percolation model for obstacles with occupation equal to $p$ and respectively the square bond percolation model for for non-cut paths with probability of $1-p$, see SI).}
 	\label{FIG1}
 \end{figure}

\textit{Model:}
Random walk models have been successfully used to describe dynamics of a wide range of biological objects ranging from animal movement, dynamics of micro-organisms to diffusion of biomolecules\cite{codling2008random}. Generally, mean square displacement, $\langle r^2 (t) \rangle$, for a randomly walking object can be written as: 
   \begin{equation}
     \langle r^2 (t) \rangle = 4 D t^\alpha +   \textbf{v} ^2t^2
     \label{eq:1}
     \end{equation}
in which $D$ is the diffusion constant and $\textbf{v}$ is the drift velocity and if $\textbf{v}=0$ walk is simple random walk (SRW). In the presence of obstacles, diffusion constant changes from a fixed parameter to a location dependent parameter, yet the mean square displacement (MSD) and time remain linearly related (of course to a limited extent) as:
\begin{equation}
      \langle r^2 (t) \rangle = 4 D_{eff} t^\alpha
      \label{eq:3}
  \end{equation}   
 in which $D_{eff}$ is the effective diffusion constant for the medium. As we will show in this paper, $D_{eff}$ plays essential role in the time scale of the events for simple random motion but not for directed motion. Without losing generality, we consider a $2D$ lattice in which a walker is located at the center of a unit as its initial location and can jump to the centers of the nearest neighbor units. In this regard, jumping length would be equal to the distance between unit centers (or edge of each unit), $\delta$, at each  time step($\tau$). In order to model environmental constraints, we consider each edge of these units as a potential obstacle. Thus, for each $(x,y)$ point we define two parameter $E(x^{+},y)$ and $E(x,y^{+})$ for corresponding edges of each unit. When there is an obstacle ($E=1$ or lines in Figure \ref{FIG1}) between two neighboring units, walker is not allowed to pass. Otherwise ($E=0$, free space in figure \ref{FIG1}), it is free to pass. As an initial condition, we screen all edges in the lattice and will consider each one as a solid line ($E=1$) with probability of $p$ and consequently no obstacle, $E=0$, with probability of $1-p$ (This process creates the square bond percolation model \cite{saberi2015recent} for obstacles with occupancy equal to $p$ and respectively the square bond percolation model for non-cut paths with occupancy of $1-p$ , see SI).
 
Generally, the Fokker-Plank equation for this model would be like below \cite{codling2008random}:
   \begin{equation}
   \frac{\partial P(x,y,t)}{\partial t}=-\vec{\nabla}.(\textbf{v}P(x,y,t))  +\vec{\nabla}.(   \textbf{D}   \vec{\nabla}P(x,y,t))
   \label{eq:4}
   \end{equation}
     in which  
$   \textbf{v} (x,y)= 
   \begin{bmatrix}
       b_{1}(x,y)  \\
       b_{2} (x,y) \\   
   \end{bmatrix} 
$ 
and 
$$ \ \textbf{D} (x,y)= 
   \begin{bmatrix}
       a_{11}(x,y) & 0  \\
       0 & a_{22} (x,y) \\   
   \end{bmatrix} 
   =
      \begin{bmatrix}
          D_{x}(x,y) & 0  \\
          0 & D_{y} (x,y) \\   
      \end{bmatrix} 
$$ with $b_{1}= \delta [r(x,y)-l(x,y)] / \tau, \text{ } b_{2}=\delta [u(x,y)-d(x,y)] / \tau $ \\  $ a_{11}(x,y)=\delta^2[r(x,y)+l(x,y)] / 2\tau $,    $ a_{22}(x,y)=\delta^2 [u(x,y)+d(x,y)] / 2\tau  $ 
where $r(x,y)$, $l(x,y)$, $u(x,y)$ and $d(x,y)$ are the probabilities of going right, left, up and the down for a single walker and for a SRW we have $r=l=u=d= 1/4$ (for more details see SI).  
 
\textit{Results:}
To study the effect of environmental obstacles (which in our model are solid lines and their number is proportional to the value of $p$) on migration of walkers, we \textbf{simulate  the motion of $10^{5}$ walkers separately moving on bond percolation lattice based on equation \ref{eq:4} with $\textbf{v}=0$.}  For $\textbf{ v} =0$ and different values of $p$ we calculated the mean square displacement($MSD$ or $\langle r^2 (t) \rangle $) by averaging over trajectories of all independent walkers in a $600\times600$ lattice and for $600$ time steps (probability of reaching to the border of lattice in 600 steps is quite small and we could be sure that none of the searching walkers would reach the border). 
 
\begin{figure} [htb]
 	\centerline{
 	\includegraphics[width=1.0 \linewidth]{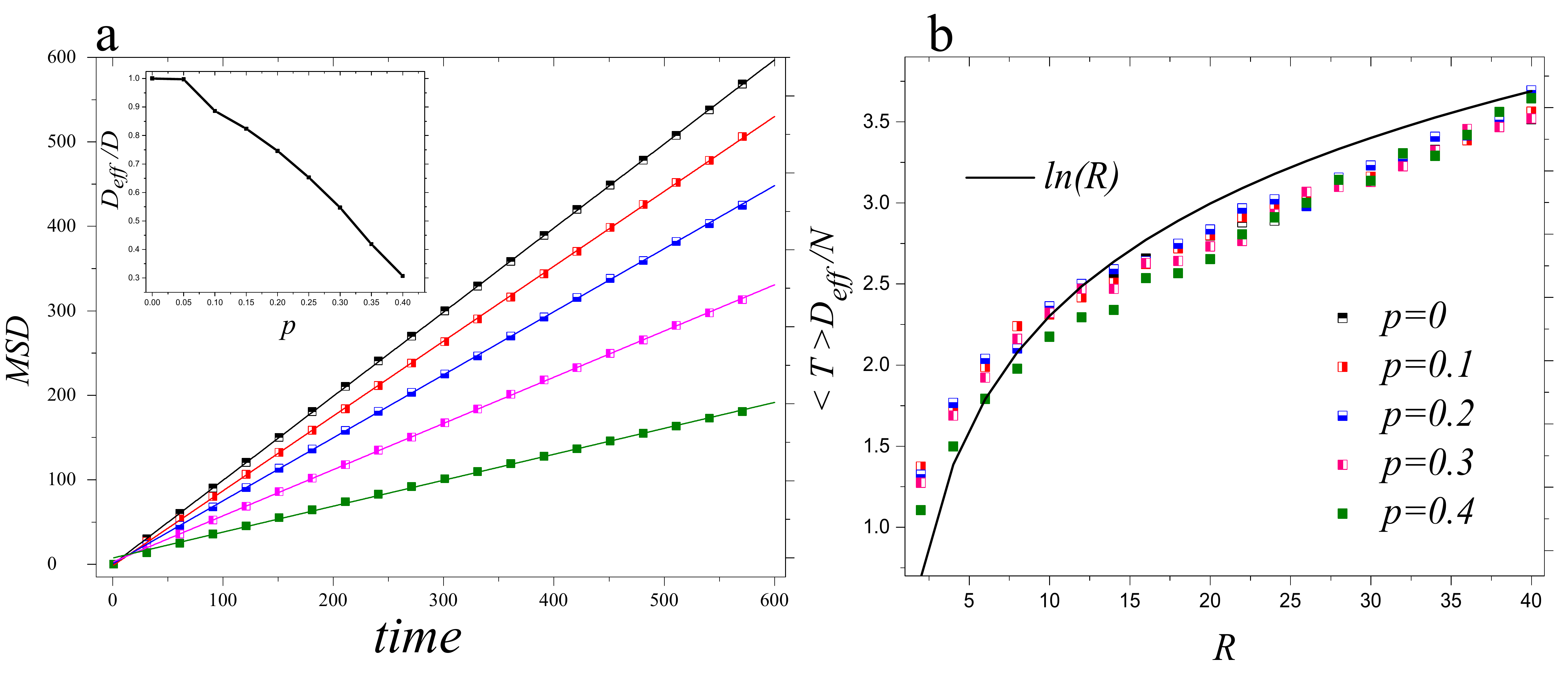}}
 	\caption{Color Online. a) Evolution of MSD per time for different values of $p$ in units of jumping length, $\delta$, and time step, $\tau$. Linear behavior which remains for $p<p_{c}=0.5$, indicates diffusion-like motion for walkers and the slopes of fitted lines are equal to $4D_{eff}$ for each $p$. As $p$ increases, $D_{eff}$ decreases.	b) Normalized MFPT, $\langle T \rangle D_{eff}/N$, per distances between the initial place of walkers and their target, $R$, for different values of $p$ fits to the universal digram of $ln(R)$. \textbf{In agreement with previous results  \cite{benichou2015mean},} this figure shows that $D_{eff}$ regulates the time scale of search. SEMs for all cases are smaller than symbol size.}
 	\label{FIG2}
\end{figure}

We first focus on search efficiency of simple random walkers. We can readily extract the main macroscopic feature of simple random walk, i.e. diffusion constant $D$, through analysis of $\langle r^2 (t) \rangle $. When $p$ increases, searching walkers will migrate slower but we define an effective diffusion constant, $D_{eff}$, as the indication of average diffusion constant for the whole medium. Using variation of MSD per time and $\langle r^2 (t) \rangle = 4 D_{eff}t $, we calculate the value of $D_{eff}$ for different values of $p$ (Figure \ref{FIG2}a). To analyze the search efficiency of random walkers, we need to calculate FPT ($T$) and its average, MFPT ($\langle T \rangle$). To do so, we simulate the \textbf{random walk process on relative percolation lattice for $10^{5}$ independent walkers and} find the time \textbf{, $T$,} that takes for \textbf{every} searching walker to reach the target. To be consistent with previous studies and to create comparable results, we work with normalized MFPT, $\langle T \rangle/N$ in which $N$ is the number of sites which the walker can visit (for distribution of $T$ see SI). In agreement with previous results \cite{benichou2015mean}, the value of $\langle T \rangle D_{eff}/N$ fits nicely the universal curve of $ln(R)$ in which $R$ is the distance between initial place of searching walkers and their target(Figure \ref{FIG2}b). 

\begin{figure}  [htb]
  	\centerline{\includegraphics[width=1.0 \linewidth]{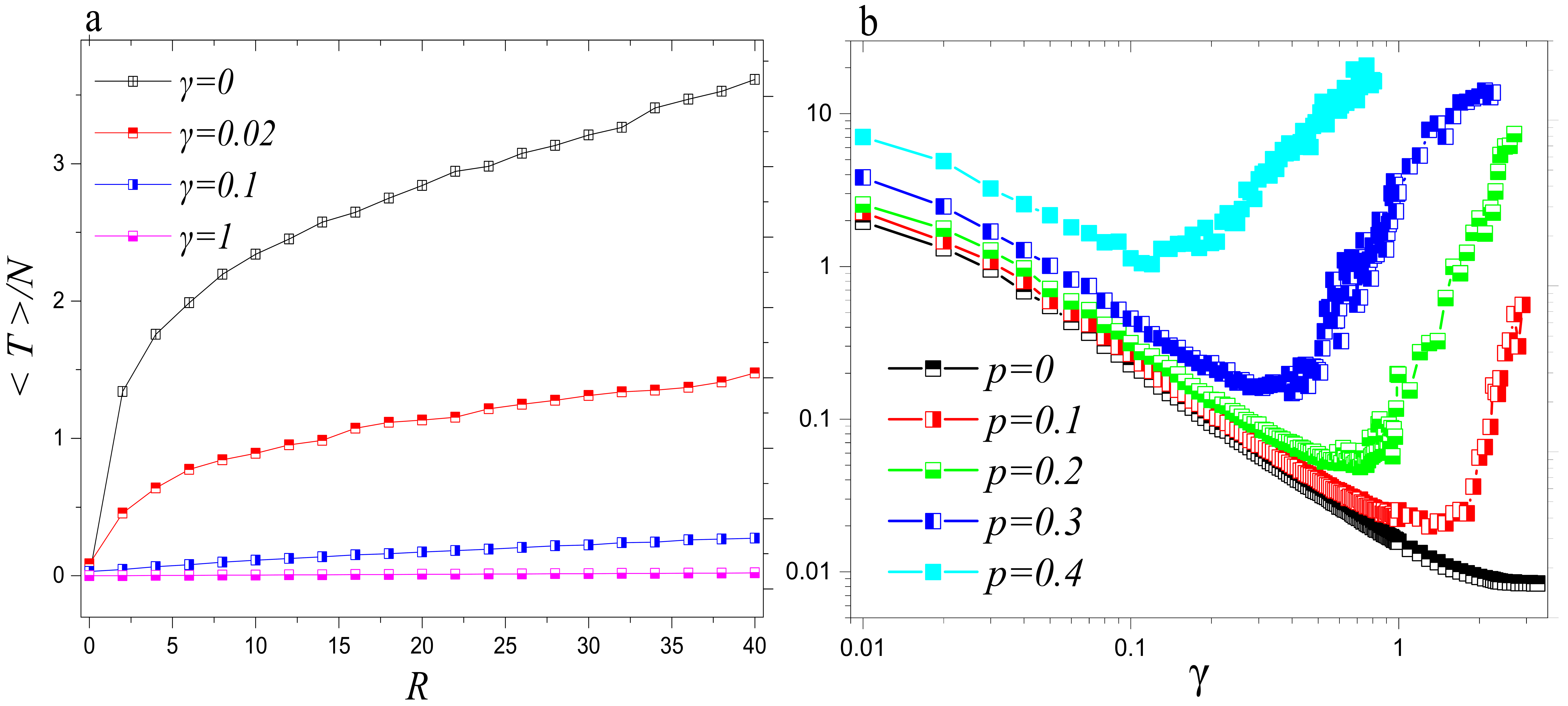}}
  	\caption{Color Online. a) Variation of $\langle T \rangle/N$ per $R$ for $p=0$ and different values of $\gamma$. As  $\gamma$ increases, $\langle T \rangle/N$ changes from $ln(R)$ for a random walker to $R/v$ for ballistic motion in which $v$ is the migration velocity.  
  	b) Changes of $\langle T \rangle/N$ per $\gamma$ for $R=30$ and different values of $p$. As $p$ increases, $\langle T \rangle/N$ deviates from the univocal behavior. For larger values of $p$, there is a $\gamma^{*}$ at which walkers have the minimum value for $\langle T \rangle$. For $p=0.4$,  $p=0.3$,  $p=0.2$ and  $p=0.1$; $\gamma^{*}=0.12$,  $\gamma^{*}=0.4$,  $\gamma^{*}=0.72$ and  $\gamma^{*}=1.3$ respectively are creating $\langle T \rangle ^{*} /N $. SEMs for all cases are smaller than symbol size.}
  	\label{FIG3}
  \end{figure}

Next we study directionally biased migration in the presence of obstacles. We consider that a walker that starts searching from its initial location $(x_{0},y_{0})$ tend to migrate toward a target at $(x',y')$.   \textbf{ Biological entities can detect the local gradients of chemicals and respond to them by moving toward the source of chemicals \cite{gire2016mice,  gaudry2012smelling, levchenko2002models, stocker2012marine, berdahl2013emergent}. Based on the Keller-Segel model \cite{keller1971model}, this process can be modeled through choosing the direction towards the target with a higher probability (see SI). In such a case} the probability of moving in the right direction at any $(x,y)$ position \textbf{can be} defined as below (other probabilities also change the same) \textit{which can be higher or lower than other probabilities based on the the values of $x$ and $x'$.}: 
 \begin{equation}
 r(x,y)  = \frac{  1-E(x^{+} ,y)}{S(x,y)}  \frac{e^{\gamma |x-x'|}}{e^{ \gamma|x-\delta-x' |}} 
 \label{eq:5}
 \end{equation}  
 
 in which $\gamma$ is the strength of migration towards the target \textit{ and is formed of ability of entities to detect and respond to chemicals gradient}. $\gamma=0$ indicates SRW (for $l$, $\delta$ changes to $-\delta$ and for $u$ and $d$, $y$ takes the place of $x$) and $S(x,y)$ is again the normalization factor which ensures $r+l+u+d=1$. 
    For $p=0$ we have  
    $b_{1}=\pm b_{2}=\mp   \delta (e^{\gamma \delta}-e^{- \gamma \delta})/\tau S(x,y)$       and  
    $ a_{11}(x,y)=a_{22}(x,y)= \delta^2  (e^{\gamma \delta}+e^{- \gamma \delta} ) / \tau S(x,y)$. 
  	 When $\gamma$ approaches $3.5$, migration changes to an entirely directed (ballistic) motion towards the target (Figure \ref{FIG3}a). To understand the effect of directionality of migration on search efficiency, we analyzed  $\langle T \rangle$ per $R$ for different values of directionality, $\gamma$. When we increase the directionality of motion, behavior of $\langle T \rangle/N$ per $R$ changes from $ln(R)$ for SRW to $R/v$ for ballistic motion in which $v$ is the migration velocity.
 
As Figure \ref{FIG3}b indicates, for $p=0$ and a fixed value of $R$, MFPT is a univocally decreasing function of $\gamma$ which means that directionality of motion always increases the search efficiency and for $\gamma>3$ $\langle T \rangle /N$ reaches to its minimum,  $\langle T \rangle ^{*} /N$, which is the time needed for directly moving from initial place towards target divided by the factor $N$. While for a non-zero $p$, the behavior of $\langle T \rangle/N$ per $\gamma$ is not univocal (Figure \ref{FIG3}b). For each non-zero value of $p$, there is a $\gamma^{*}$ that creates the minimum value of $\langle T\rangle/N$ which identifies the optimized search strategy. This result implies that in a heterogeneous medium, a purely random or purely directed migration towards target is not the optimized strategy. Instead, a combination of directed and random motion is the optimal choice and the density of obstacles determines the contributions of each mode.  
 
Next we ask how target frequencies and motility of targets affect search efficiency.
Number of randomly distributed targets within medium, $n$, regulates the value of $\langle T \rangle/N $ through decreasing the distance in which we expect the searching walker to see a target by the factor of $\sqrt{n}$ and consequently $\langle T \rangle/N $ with the factor of $n$. 
Furthermore, instead of a fixed target, we consider that at each time step, the target is able to move randomly by probability of $p_{m}$ or remains at its location with probability of $1-p_{m}$. In the targets' frame of reference, target is fixed but at each time step the walker moves  $1+p_{m}$ times. As a result, after $M$ steps it had moved $M \times (1+p_{m})$ times and $\langle T \rangle /N$ should decrease by the factor of $(1+p_{m})$.

\begin{figure} [htb]
 \centerline{\includegraphics[width=1.0 \linewidth]{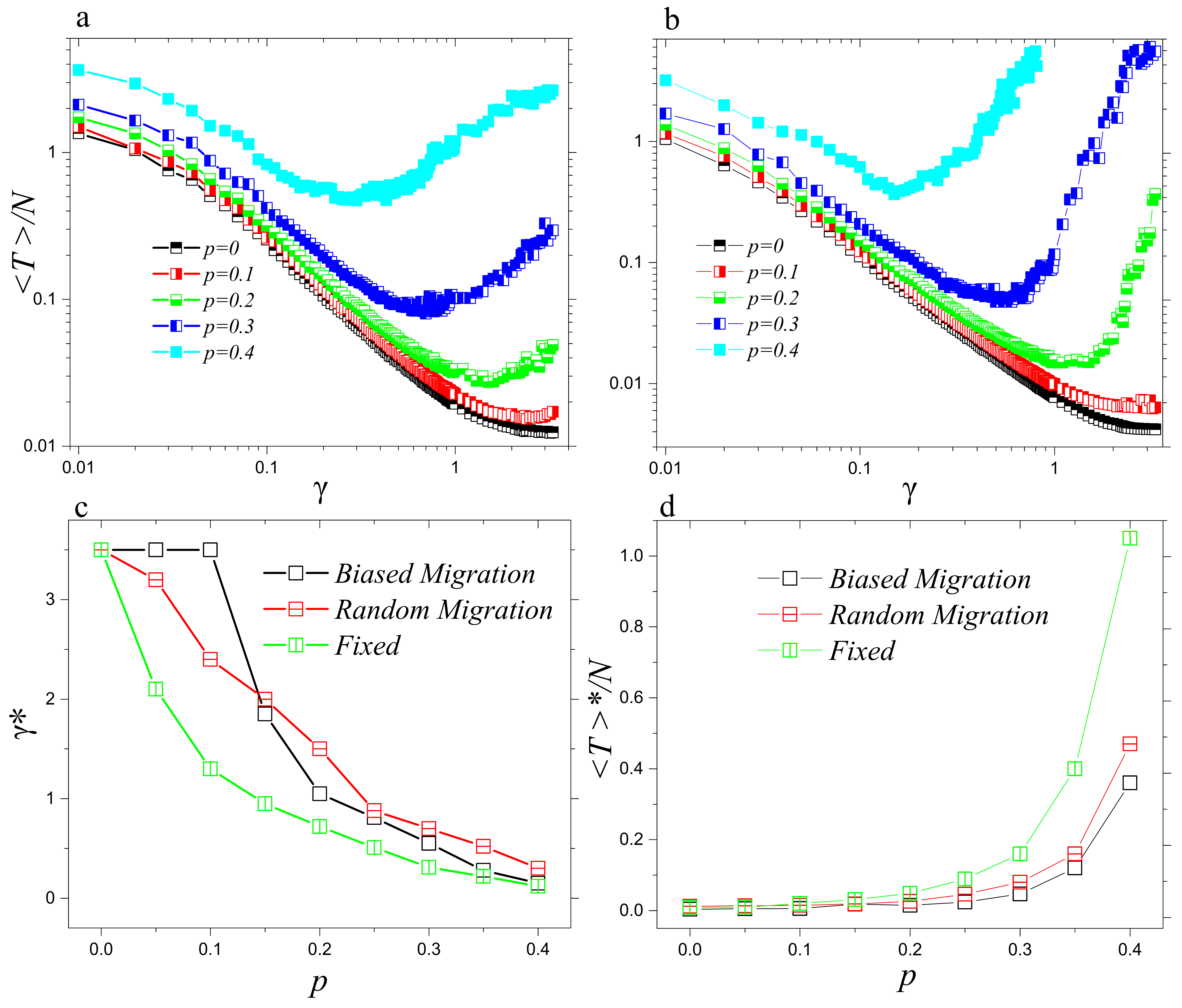}}
 \caption{Color Online. a) Behavior of $\langle T \rangle /N$ per $\gamma$ when walker migrate towards the target with directionality of $\gamma$ and the target itself moves randomly.
 b) Behavior of $\langle T \rangle /N$ per $\gamma$ when both target and walker migrate towards each other with directionality of $\gamma$. 
 c)Behavior of $\gamma ^{*}$ as the $\gamma$ which minimizes $\langle T \rangle /N$ and consequently  generates the optimized search strategy for all studied cases.
  d)Behavior of  $\langle T \rangle ^{*}  /N$ as the minimum value of $\langle T \rangle /N$ for all studied cases.
 SEMs for all cases are smaller than symbol size, see SI.}
 \label{FIG4} 
\end{figure}

 As another possibility, we considered a walker chasing a target which in turn moves randomly. Effect of directionality of motion and density of obstacles were studied (Figure \ref{FIG4}a). Finally, it may happen that both targets and walkers search for each-other according to equation \ref{eq:5} (Figure \ref{FIG4}b).  Migration of target (both randomly and towards searching walkers) does not change the general behavior of $\langle T \rangle /N $ for different values of $p$ but the value of $\gamma $  for corresponding optimized search strategy, $\gamma ^{*}$ would be different (Figure \ref{FIG4}c). Dynamics of $\langle T \rangle ^{*} /N $ for all studied cases was obtained (Figure \ref{FIG4}d). 

Using MSD analysis to study random walk is not reliable for all cases \cite{textor2011defining, banigan2015heterogeneous}. For small values of $\gamma$, MSD does not change whereas $\langle T \rangle$ might changed dramatically. Besides, for larger values of $\gamma$ when we increase the value of $p$, analysis of MSD shows that a super-diffusive motion is gradually converted to a diffusive motion and then to a sub-diffusive motion.
     

\textbf{The results of our study can be generically applied to a wide range of problems including cell migration, animal movements and dynamics of non-biological microparticles. For example, as shown In the supplementary information, our findings can be applied to shed light on some of current open problems related to T cell migration. }

\textit{ Discussion:} 
Here we simulated a wide range of migration dynamics from random migration to a pure directed migration towards targets and studied the unexplored effect of directional bias on $\langle T \rangle $ and density of obstacles and motility of targets. Expectedly, we find that \textbf{density of obstacles dose not change the} $\langle T \rangle $ for simple random walk. For a directionally biased migration, directionality always decrease $\langle T \rangle $ in the absence of obstacle, and the minimum value of $\langle T \rangle $ as the indicative of the optimized migration strategy corresponds to the largest directionality,\textbf{ $\gamma^{*} \sim 3.5$} (ballistic motion). In the presence of obstacles, instead of a pure directed motion, a combined motion \textbf{, $0<\gamma^{*} < 3.5$,} leads to the minimum value of $\langle T \rangle $ and thus the optimized search strategy. The physical structure of environment thus dramatically affect search efficiency and randomness plays a beneficial role in finding targets. When the target frequency increases or the targets are able to migrate (both randomly and directed towards walker), they are easier to find (not for higher values of $p$). In the presence of obstacles, directionality plays the same dual role.  Finally, physical obstacles could change the dynamics of motion and convert a super-diffusive motion to a sub-diffusive one.
 
\textit{  Acknowledgments}: 
We acknowledge A. Gerard and M. Javadi for reading our manuscript and their comments. We also thank Department of Physics, Tehran University for the computational facility.

\bibliography{draft}
\end{document}